%% file: main.tex
\documentclass[sigconf, nonacm]{acmart}
\usepackage{float}
\usepackage{graphicx}
\usepackage{caption}
\usepackage{subcaption}
\usepackage{placeins}
\AtBeginDocument{%
  \providecommand\BibTeX{{%
    \normalfont B\kern-0.5em{\scshape i\kern-0.25em b}\kern-0.8em\TeX}}}

\begin{document}

\title{Studying the Potential of Automatic Optimizations in the Intel FPGA SDK for OpenCL}

\author{Adel Ejjeh}
\email{aejjeh@illinois.edu}
\affiliation{
\institution{University of Illinois at Urbana-Champaign}
\city{Urbana}
\state{IL}
}
\author{Vikram Adve}
\email{vadve@illinois.edu}
\affiliation{
\institution{University of Illinois at Urbana-Champaign}
\city{Urbana}
\state{IL}
}
\author{Rob A Rutenbar}
\email{rutenbar@pitt.edu}
\affiliation{
\institution{University of Pittsburgh}
\city{Pittsburgh}
\state{PA}}
\affiliation{
\institution{University of Illinois at Urbana-Champaign}
\city{Urbana}
\state{IL}
}
\vspace{-12pt}
\begin{abstract}
High Level Synthesis (HLS) tools, like the Intel FPGA SDK for OpenCL, improve design productivity and enable efficient design space exploration guided by simple program directives (pragmas), but may sometimes miss important optimizations necessary for high performance. In this paper, we present a study of the tradeoffs in HLS optimizations, and the potential of a modern HLS tool in automatically optimizing an application. We perform the study on a 5-stage camera ISP pipeline using the Intel FPGA SDK for OpenCL and an Arria 10 FPGA Dev Kit. We show that automatic optimizations in the HLS tool are valuable, achieving a up to 2.7$\times$ speedup over equivalent CPU execution. With further hand tuning, however, we can achieve up to 36.5$\times$ speedup over CPU. We draw several specific lessons about the effectiveness of automatic optimizations guided by simple directives, and the nature of manual rewriting required for high performance.

\end{abstract}

%
%

\maketitle
\input{Intro.tex}
\input{Background.tex}
\input{Optimizing.tex}
\input{Results.tex}

\input{Related.tex}
\input{Conclusion.tex}


 \bibliographystyle{ACM-Reference-Format}
\bibliography{references}

\end{document}

%% file: Intro.tex
\section{Introduction} \label{sec:intro}
In recent years, High-Level Synthesis (HLS) has gained a lot of traction in the accelerator design community as a faster means of designing high-performance accelerators on Field Programmable Gate Arrays (FPGAs) and Application Specific Integrated Circuits (ASICs) \cite{VivadoHLS,SDAccel,zhang2008autopilot,IntelFPGAOpenCL,czajkowski2012opencl,maxcompiler,koeplinger2018spatial,pu2017programming, canis2011legup, canis2013legup, hegarty2014darkroom, wei2013improving}. With the right optimizations and tuning, HLS allows designers to reach the same end goal as Hardware Descriptive Languages (HDLs) like Verilog and VHDL, in a fraction of the design time with minimal loss of performance. In fact, recent literature has been showing a wide adoption of HLS in designing accelerators for new applications in a multitude of domains \cite{nakahara2018lightweight,Cabal:2018:CFP:3174243.3174250,zohouri2018combined}. 

A successful example of an HLS system is AOC, the Intel FPGA SDK for OpenCL \cite{IntelFPGAOpenCL}. AOC allows designers to use OpenCL for designing their accelerator while targeting Intel's family of FPGAs. However, getting good performance with AOC is non-trivial, requiring extensive tuning. Intel has released two supporting documents, a Programming Guide~\cite{AOCLProgrammingGuide} and a Best Practices Guide~\cite{AOCLBestPractices} to provide designers with optimization techniques for achieving the best possible performance with AOC. In addition, AOC provides two reporting mechanisms: 1) a pre-synthesis Optimization Report that gives the designer a quick estimate of performance, allowing for rapid code modification without having to go through synthesis, and 2) a Profile Report that gives profiling information of the kernel physically running on the FPGA (post synthesis and place-and-route).

However, even with the guidance that Intel provides in its manuals, getting the best performance possible still presents at least three challenges. First, the programmer must often use compiler directives (in the form of pragmas) to provide information that may not be tractable for the compiler to prove automatically, such as for interprocedural pointer aliasing or loop-carried dependences. Second, it is often the case that different optimizations come with different tradeoffs. Navigating the design space of these optimizations requires the programmer to go through many iterations of their designs, many of which need to go through the complete synthesis flow, because the Optimization Report is not always indicative of the actual performance in hardware. Third, and most problematic, some important optimizations may be lacking in the compiler entirely (for a variety of reasons), and programmers may need to make significant manual changes to the code in order to accomplish those optimizations.

In this work, we study the tradeoffs of different optimization techniques, and the potential of automatic directive-driven optimizatons in AOC, on a 5-stage Image Signal Processing (ISP) pipeline that takes a raw image produced by camera sensors, and converts it into a viewable image format. Our target device is an Intel Arria 10 GX FPGA Development Kit which carries a GX 1150 FPGA. We show through our study that while some optimizations have clear-cut benefits and always yield improvements, others behave differently depending on the kernel being optimized. We also show that the amount of performance improvement that can be achieved automatically by the compiler, with only simple directives and minimal code modifications, is useful but limited, and that, in most cases, achieving the best performance requires significant modifications to the code.

In order to perform our study, we implemented the stages of the ISP pipeline as separate kernels in OpenCL using the Intel-recommended single work item model. Next, we used the pre-synthesis Optimization Report to choose different optimizations from Intel's Best Practices Guide that would help us achieve perfect pipelining of the outer loops in the kernels. We profiled the resulting design and found that significant performance improvements were possible but not achievable using automated optimizations implemented in the compiler (even with programmer directives), and therefore required further hand-tuned optimzation of the kernels. While some of these optimizations are described in the Best Practices Guide (like buffering of inputs), a very important transformation (similar to unroll-and-jam~\cite{callahan1988estimating}) is not. We performed both kinds of optimizations through manual hand-tuning of the code. Our final results show that for the full pipeline, automatic optimizations give a 3.3$\times$ slowdown, while hand tuning the kernels gives us up to 4.6$\times$ speedup compared to CPU execution. Looking at individual kernel performance, we find that automatic optimizations can achieve up to 2.72$\times$ speedup, while hand tuning the kernels allows us to achieve close to 36.5$\times$ speedup compared to CPU.

In our study, we treated the 5 stages of the ISP pipeline as distinct kernels, each exhibiting its own characteristics, to try and identify patterns in the effect of the optimization techniques that we apply. Through comparing the behaviors of the 5 kernels, we were able to learn the following lessons:
\begin{enumerate}
    \item Our most important finding is that automatic optimizations that are performed by AOC, even those guided by directives, are not sufficient to achieve the best performance possible without extensive manual modifications by the designer.
    \item In spite of the above limitation, using `\texttt{restrict}' and `\texttt{ivdep}' keywords almost always provide a decent boost in performance by allowing AOC to avoid memory dependence assumptions.
    \item Rewriting the kernels to increase the amount of independent operations within the body of a pipelined outerloop provides significant performance benefits by increasing spatial parallelism.
    \item If read-only data fits on the FPGA, manually buffering the data gives better performance than depending on AOC's constant memory, in all but one case, since it removes the overhead of the cache structure.
    \item Finally, if inner loops are present in the body of the pipelined loop, even partially unrolling them gives a boost in performance since it increases the amount of parallel operations.
\end{enumerate}

By performing this study, we identified the limitations of automatic optimization in a modern commercial HLS tool like AOC, which motivates the need for more extensive compiler optimizations in these tools. This is an area of research that we are interested in and are pursuing as future work.

The rest of the paper is organized as follows: Section \ref{sec:background} provides a background about the ISP pipeline and the Intel FPGA SDK for OpenCL, followed by the details of how the optimizations we used, and how we applied them to each kernel in Section \ref{sec:opt}. Next, Section \ref{sec:eval} provides our experimental evaluation, followed by some of the related work in Section \ref{sec:related}, after which we conclude in Section \ref{sec:conclusion}.

%% file: Background.tex
\section{Background} \label{sec:background}
\begin{figure*}[!htb]
    \centering
    \includegraphics[width=\textwidth]{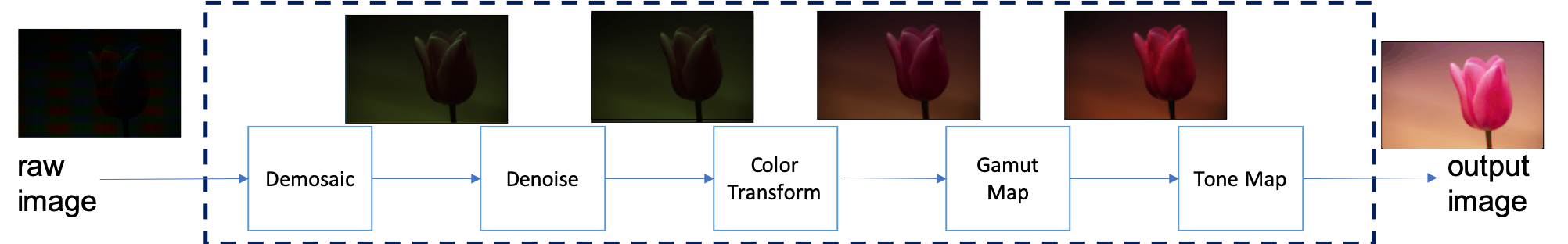}
    \caption{Architecture of the 5-stage camera ISP pipeline.}
    \label{fig:cava}
\end{figure*}

\subsection{The 5-Stage Camera ISP Pipeline} \label{ssoc:cava}
To perform our case study, we use the 5-stage Camera ISP pipeline that is shown in Figure \ref{fig:cava}~\cite{yaoyuannnn}. It takes in a raw image that is produced by camera sensors, and generates a useful image that can be displayed. The main stages of the pipeline are described below:

\textit{Demosaic:} This stage applies a Bayer Filter color filter array on each raw pixel to interpolate it's true R-G-B values. It produces a mosaic of RGB pixel intensities. \cite{wiki:demosaic}

\textit{Denoise:} This stage applies a local nonlinear interpolation denoising algorithm to reduce the level of noise in the image. \cite{wiki:denoise}

\textit{Color Space Transform / White Balancing:} This stage performs color balancing by multiplying the RGB color value at each point with a 3x3 diagonal matrix whose values are configurable. It preserves the neutrality of neutral colors. \cite{wiki:transform}

\textit{Gamut Mapping:} This stage maps the colors of the original image to a set of restricted available colors of an output device without compromising the original image. To do so, it first computes the L2-norm from each pixel to the set of control points that represent the target gamut. Then, the L2 distances are weighted and summed. Finally, a bias is added to implement a radial basis function. This is the most computationally-intensive kernel and serves as the bottleneck for the pipeline. \cite{wiki:gamut}

\textit{Tone Mapping:} This stage approximates images with a higher dynamic range than the output device. This is done by using a Tone Map Operator to squeeze the original dynamic range of the image into the lower range of the output device. \cite{wiki:tone}

\textit{Profile of the pipeline:} The execution time breakdown of the pipeline is as follows: 99\% in Gamut Map, 0.8\% in Denoise, 0.04\% in Transform, 0.03\% in Demosaic, and 0.02\% in Tone Map. Therefore, we can see that Gamut Map significantly outweighs all the other kernels, and we focus our efforts on achieving the best speedup for this kernel, while still studying the tradeoffs in applying different optimizations on the other kernels.

\subsection{Intel FPGA SDK for OpenCL} \label{ssec:aocl}
The Intel FPGA SDK for OpenCL is an HLS tool that allows hardware designers to use OpenCL, instead of HDLs, for programming Intel FPGAs. The main tool in the SDK is the Altera OpenCL Compiler (AOC), an HLS compiler that compiles the OpenCL kernels into RTL, then runs them through Intel Quartus to synthesize them and generate an FPGA bitstream. AOC's strong suit is in automatically pipelining loops in a kernel and trying to achieve perfect pipelining with an initiation interval (II) equal to 1. As such, it is recommended that designers implement their kernels as Single Work Item Kernels (SWIK), as opposed to the traditional multithreaded kernels that OpenCL is known for, in order to maximize AOC's ability to pipeline the kernel's execution, and extract as much parallelism as possible. To facilitate the process of optimizing kernels using Intel's OpenCL SDK, AOC provides three key resources: 1) An Optimization Report, which can be generated through a quick intermediate compilation step, that contains information about estimated resource utilization and the status of all the loops (unrolled, pipelined -- along with the estimated II). 2) A Profile Report, which requires a full synthesis and place-and-route to be completed, and provides profiling information related to the kernel performance (operating frequency, execution time, memory bandwidth, etc...). 3) A Best Practices Guide, which is a document that provides different optimization techniques to assist the designer in selecting optimization decisions based on the results of the optimization and profiling reports. Next, we will describe the main information that is reported by the Optimization Report and Profiling Report.

\textit{Understanding the Optimization Report:}
The Optimization Report provides an analysis of the loops in the kernel, identifying which loops were unrolled, and which were pipelined along with their corresponding II. If the II > 1, the report provides an explanation of what might be the bottlenecks that prevented perfect pipelining of the loops, along with pointers to the corresponding sections in the Intel guides that might provide techniques to improve the II. In addition to loop information, the report provides a ``System Viewer''. This viewer shows the basic blocks of the kernel and provides information about the start cycle, end cycle, and latency of each basic block, along with the structure of local memories.

\textit{Understanding the Profile Report:}
The Profile Report is an essential tool for understanding the performance of the generated kernel. It provides the total execution time of the kernel, the operating frequency, and the global bandwidth to DRAM. It also provides details about each load and store in the kernel.

%% file: Optimizing.tex
\section{Optimizing the ISP pipeline} \label{sec:opt}
In this section, we will first describe our baseline implementation of the pipeline, followed by a description of the different optimizations that we applied along with our optimization strategy.

\subsection{Baseline} \label{ssec:baseline}
For our baseline, we use a simple `single work item' OpenCL implementation that is a direct mapping of the original C implementation of the pipeline into OpenCL. In this implementation, all arrays are stored in OpenCL global memory (FPGA on-board memory), and kernels communicate through global memory as well -- each kernel writes its output to global memory, which is then read by the next kernel. Our pipeline implimentation operates on images with three channels (`R', `G', and `B'). Images are stored in memory in row-major order such that all the rows of the `R' channel come first, followed by the rows of the `G' channel, which are then followed by the rows of the `B' channel. In the baseline implementation, all the kernels, except Demosaic, operate on the 3 channels sequentially, and produce all the pixels of `R' followed by `G', then `B'.

\subsection{Automatic Optimizations} \label{ssec:autoopt}
Some of the optimization techniques that are provided by Intel's guides, and that we use in this study, require minimal code modifications in the form of directives, and rely on the compiler to automatically optimize the kernel based on the information that these directives provide. We will now describe the three optimizations that we use:
\begin{itemize}
    \item \textit{The ``\texttt{restrict}'' keyword:} Is an attribute that can be used to mark the different pointer operands of a kernel as non-aliasing. This allows AOC to avoid making conservative assumptions about whether or not the operands may alias.

    \item \textit{The ``\texttt{ivdep}'' pragma:} Is a directive that can be used to instruct AOC to ignore any assumed loop-carried dependecies that enforce serialization of loop iterations, thus limiting the ability of AOC to pipeline the loops.
    
    \item \textit{The ``\texttt{constant}'' attribute:} This attribute can be added to pointer operands of the kernel that are read-only. It instructs AOC to cache accesses to these operands in a global Constant Memory Cache that it creates using the FPGA Block RAM. A single constant cache of configurable size -- the default of which is 16KB -- is shared among all the ``constant'' operands of all the kernels that would be running on the FPGA at the same time.
    
    \item \textit{The ``\texttt{unroll}'' pragma:} Is a directive that can be used to instruct AOC to attempt unrolling a loop. A specific unroll factor can be specified, otherwise AOC attempts to fully unroll the loop. If done on an inner loop of a pipelined outerloop, it gives AOC the potential to increase spatial parallelism by having more parallel computations to perform per iteration.

\end{itemize}

\subsection{Manual Optimizations} \label{ssec:manopt}
In addition to the automatic optimization techniques, Intel's guides recommend other optimizations that require significant code modifications, thus relying more heavily on the designer. We also combine in this category other hand-tuned optimizations that are not specifically recommended by Intel, but rather stem from the natural design considerations of spatial parallelism on the FPGA. The following describe the manual optimizations that we use in our study:
\begin{itemize}
     \item \textit{Using local memory when possible:} This represents a set of optimizations recommended by Intel that involve moving computation to local memory when possible. These techniques include copying read-only kernel inputs that possess temporal locality from global to local memory, and privatizing any kernel operands that are only generated and accessed in the kernel.
     
     \item \textit{Manual loop modifications:} There are scenarios where loop nests could be rewritten with a combination of unrolling, interchange, and fusion in order to maximize the amount of spatial parallelism by exposing more independent computations per iteration of a pipelined loop. We perform a variety of these optimizations as described Section \ref{ssec:strategy} below.

     \item \textit{Using Intel Channels:} Intel Channels are FIFO buffers that can be used to transfer data between kernels. This can be useful when multiple kernels are running on the FPGA at the same time, and they exhibit a producer-consumer behavior. By using channels, all the data can be transferred between the kernels locally on the FPGA without having to write to/read from global memory. In many cases, using channels require modifications to the kernel code to make sure that the memory access pattern of the producer and consumer match.
\end{itemize}

\subsection{Optimization Strategy} \label{ssec:strategy}
Starting with the baseline from Section \ref{ssec:baseline}, we performed different combinations of the optimizations that we discussed in Sections \ref{ssec:autoopt} and \ref{ssec:manopt} on each kernel. We carefully chose optimizations to apply based on the results of the AOC Optimization Report and the Profile Report. We break down our study of the optimizations into four steps:
\begin{enumerate}
    \item First, we apply the `\texttt{restrict}' and `\texttt{ivdep}' keywords, to achieve perfect pipelining of the outer loop, and study the extent of which AOC can automatically optimize the kernels with these keywords in place.
    \item Next, we apply manual rewrites to the code, such that each kernel generates the corresponding pixel of the `R', `G', and `B' channels in the same iteration as opposed to generating them sequentially. This step involves a combination of loop unrolling, loop interchange, and loop fusion. By performing this step, we are increasing the amount of parallel computations in a single iteration of the outerloop, which in turn should allow AOC to exploit the spatial parallelism that is available.
    \item After that, we try to further optimize the kernels by optimizing the memory accesses. In this step, we study the effects of using `constant' memory versus manually buffering for all the read-only inputs of each kernel that demonstrate temporal locality.
    \item Finally, we use the `\texttt{unroll}' pragma on any inner loops to increase the spatial parallelism that can be exploited in each iteration, and as such further improve the performance.
\end{enumerate}

Once we have the fully optimized version of each kernel, we make an additional modification for the full pipeline, where we use Intel Channels to pass data between the kernels on-chip rather than using global memory. We use this to study the effect of channels on our pipeline. We also study two other versions of the full pipeline, one having the best performing version of each kernel, and one having the best performing version of each kernel using only the autmatic optimizations that we described in Section \ref{ssec:autoopt} above.

We will now describe how each kernel in the pipeline was optimized.

\begin{figure}
    \centering
    \includegraphics[width=\columnwidth]{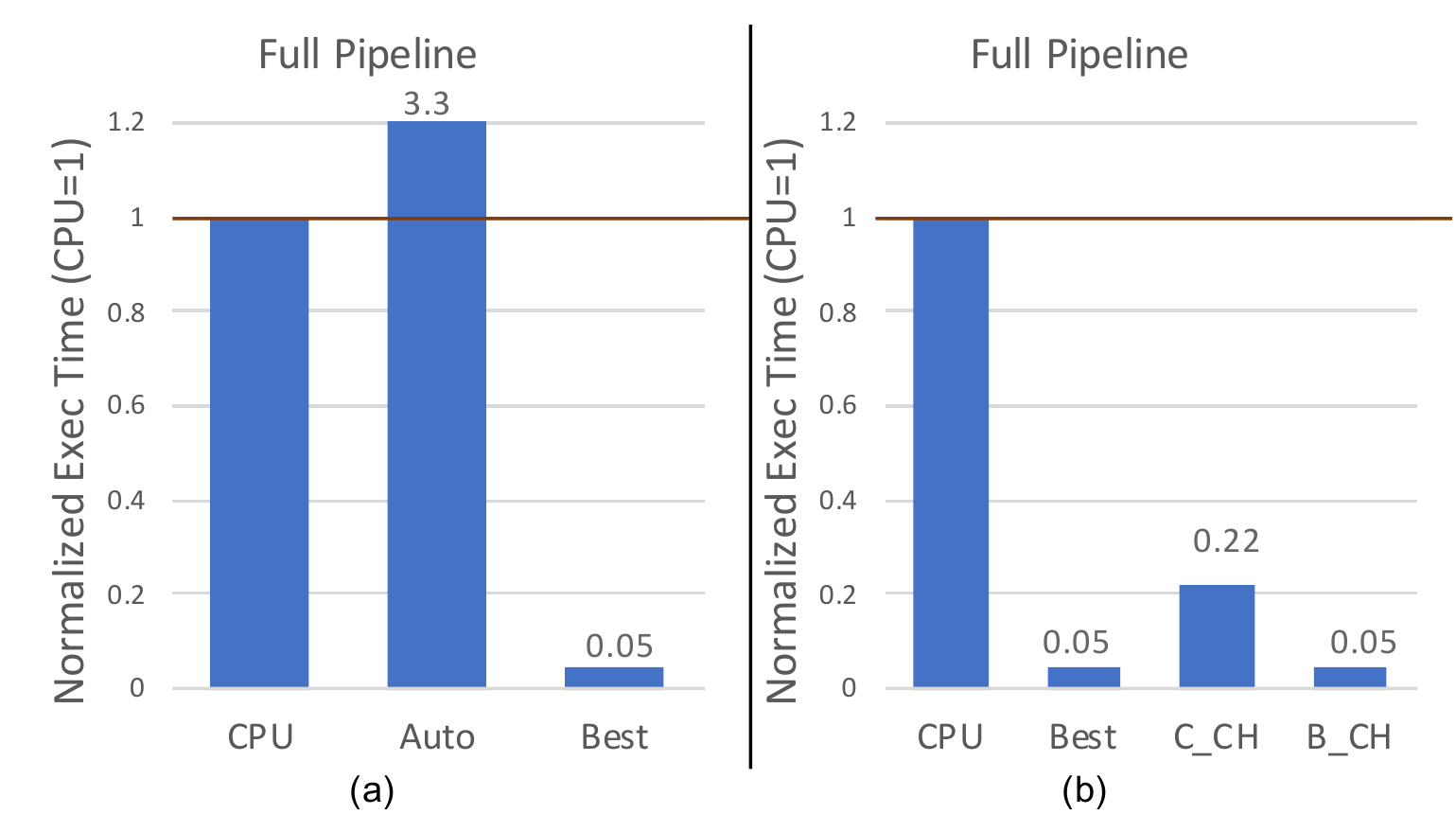}
    \caption{Full pipeline results. Best: best version of each kernel, Auto: best version of each kernel with only automatic optimizations, C\_CH: Intel channels with constant memory, B\_CH: Intel channels with manual buffering.}
    \label{fig:full}
\end{figure}

\subsubsection{Demosaic:}
This kernel has a doubly-nested loop that goes over the rows and columns of the image and calculates the `R', `G', and `B' values of each pixel, while reading the input from global memory, and writing the output to global memory. As such, we only apply step 1 from our optimziation strategy to it, since it does not have read-only operands, and it cannot benefit from rewrites.

\subsubsection{Denoise:}
This kernel has a triply-nested loop that iterates over the image channels, rows, and columns. For each input pixel (which is coming from the output of the previous kernel), it reads a 3x3 tile centered around that pixel, sorts the values, and sets the corresponding output pixel to the median of the sorted values. Optimizing the sorting algorithm is beyond the scope of this paper; as such, we can only apply steps 1 and 2 from our strategy to this kernel.

\subsubsection{Color Space Transform:}
This kernel loops over the channels, rows, and columns of the image, and for each output pixel it reads the `R', `G', and `B' values of input pixel, multiplies them by a column from a 3x3 matrix (one column for each channel), and sums up the products. The 3x3 matrix is a read-only input, as such we apply steps 1, 2, and 3 from our strategy to this kernel. We will refer to this kernel as Transform in the rest of the paper.

\subsubsection{Gamut Map:}
For each input pixel, the Gamut Map kernel calculates the L2 distance between the `R', `G', and `B' values of that pixel and each control point. Then, the L2 distances are weighted and summed. Finally, a bias is added to that sum by multiplying each `R', `G', and `B' value with a coefficient and adding them up. The final result is the corresponding output pixel. As such, the Gamut Map kernel is the most computationally intensive kernel in the pipeline. Given that the L2 distance calculation is performed in an innermost loop, we apply all 4 steps of our strategy to this kernel.

\subsubsection{Tone Map:}
The Tone Map kernel reads each input pixel, uses its value to access a \texttt{tone\_map} matrix, and sets the corresponding value from the matrix as the output pixel value. Given that this kernel has a read-only input and potential for rewrite, we apply steps 1, 2, and 3 from our strategy to it.

%% file: Results.tex
\section{Experimental Evaluation} \label{sec:eval}
In this section, we start by presenting our evaluation methodology. Next we present the results of the full pipeline, which as an overview of the optimizations. Finally, we show the results of optimizing each kernel seperately, to highlight the effect of each optimization.
\subsection{Methodology}
Our software infrastructure consists of the Intel FPGA SDK for OpenCL version 18.1 (along with Quartus Pro 18.1) running on Ubuntu 16.04 LTS. Our target device is an Intel Arria 10 GX FPGA Developtment Kit which houses an Arria 10 GX1150 FPGA. Our CPU implementation is single-threaded running on an Intel Xeon E3-1240 V2 @ 3.40GHz. The operating frequency of the kernels on the FPGA ranges between 210MHz and 360MHz.  

We synthesized different versions of each kernel, and of the entire pipeline, according to our optimization strategy described in Section \ref{sec:opt}. We tested the execution with multiple different input images of the same size, and confirmed that the results are similar. As such, we present the results for a single image, averaged over 10 runs.
\subsection{Full Pipeline Result}
\begin{figure*}[!h]
    \centering
    \includegraphics[width=\textwidth,height=4cm]{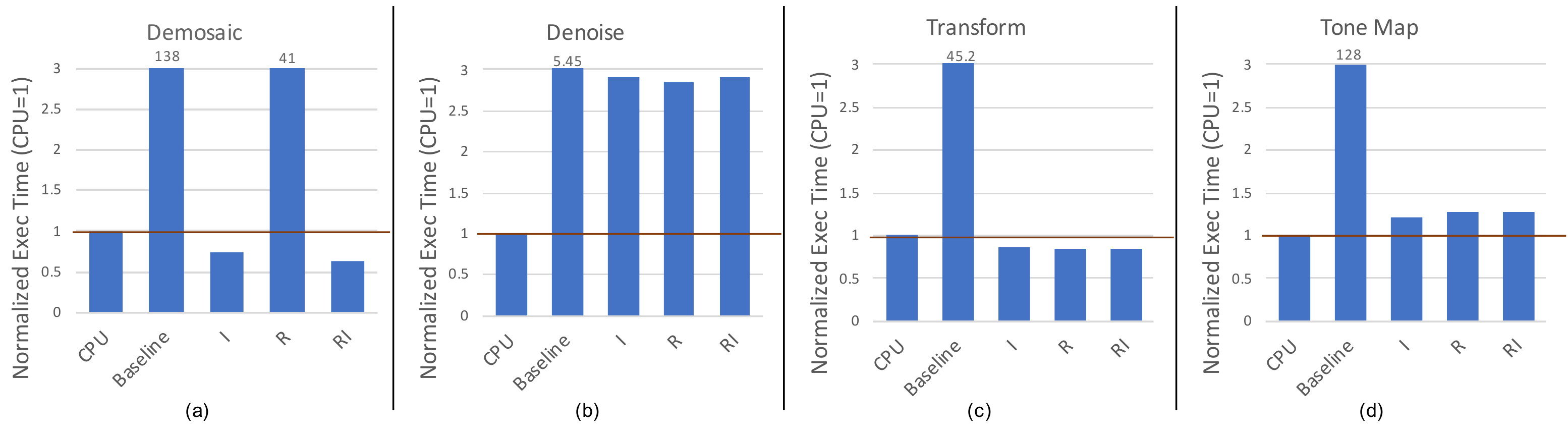}
    \caption{Effect of `\texttt{restrict}' (R) and `\texttt{ivdep}' (I) keywords on each kernel. (RI=R+I)}
    \label{fig:RI}
\end{figure*}

\begin{figure} [!h]
    \includegraphics[width=\columnwidth]{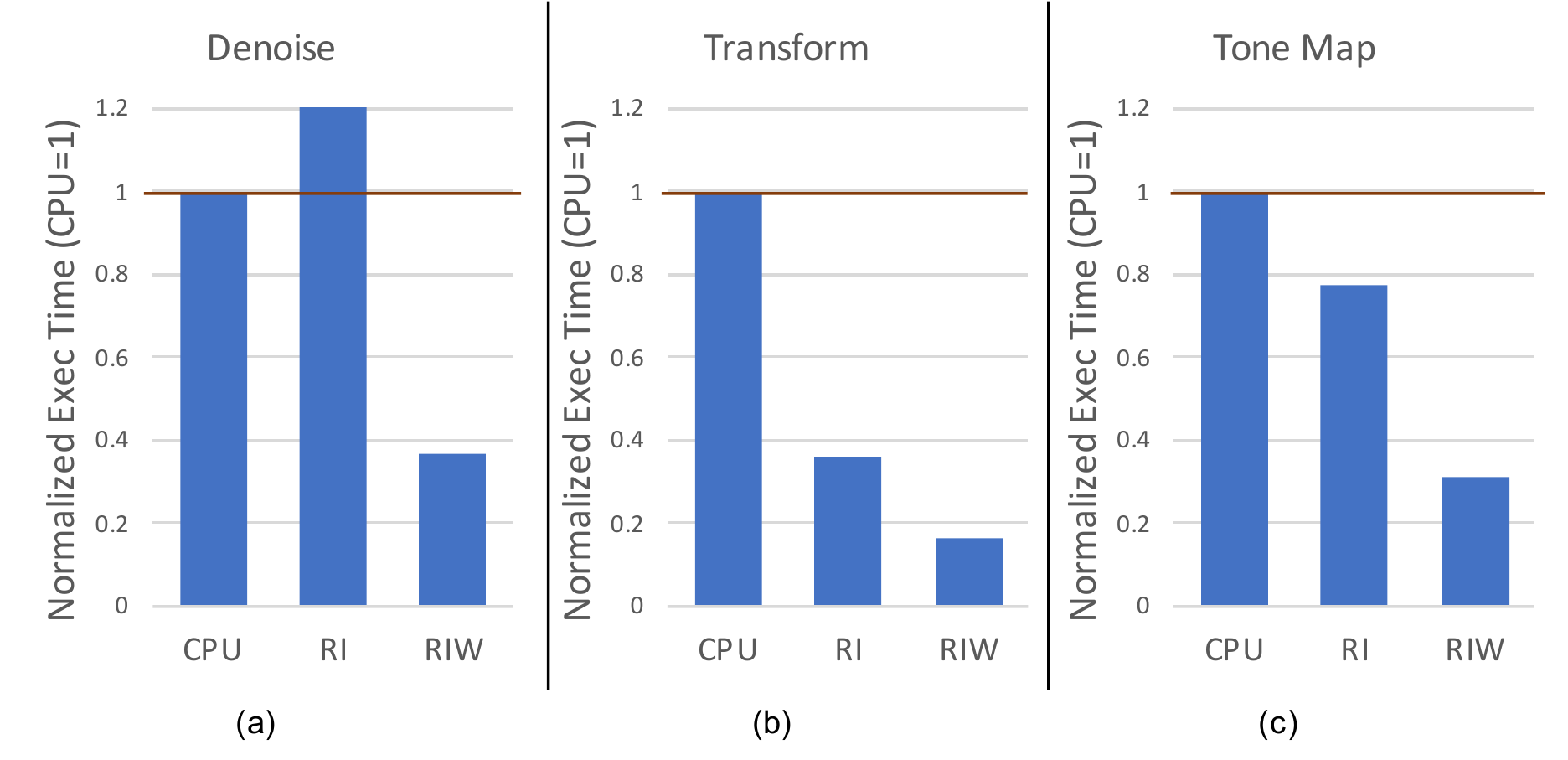}
    \caption{Effect of rewrites (W) on each kernel (see Sections \ref{ssec:manopt} and \ref{ssec:strategy}). Does not apply to Demosaic. }
    \label{fig:W}
\end{figure}

\begin{figure} [!h]
    \includegraphics[scale=0.6]{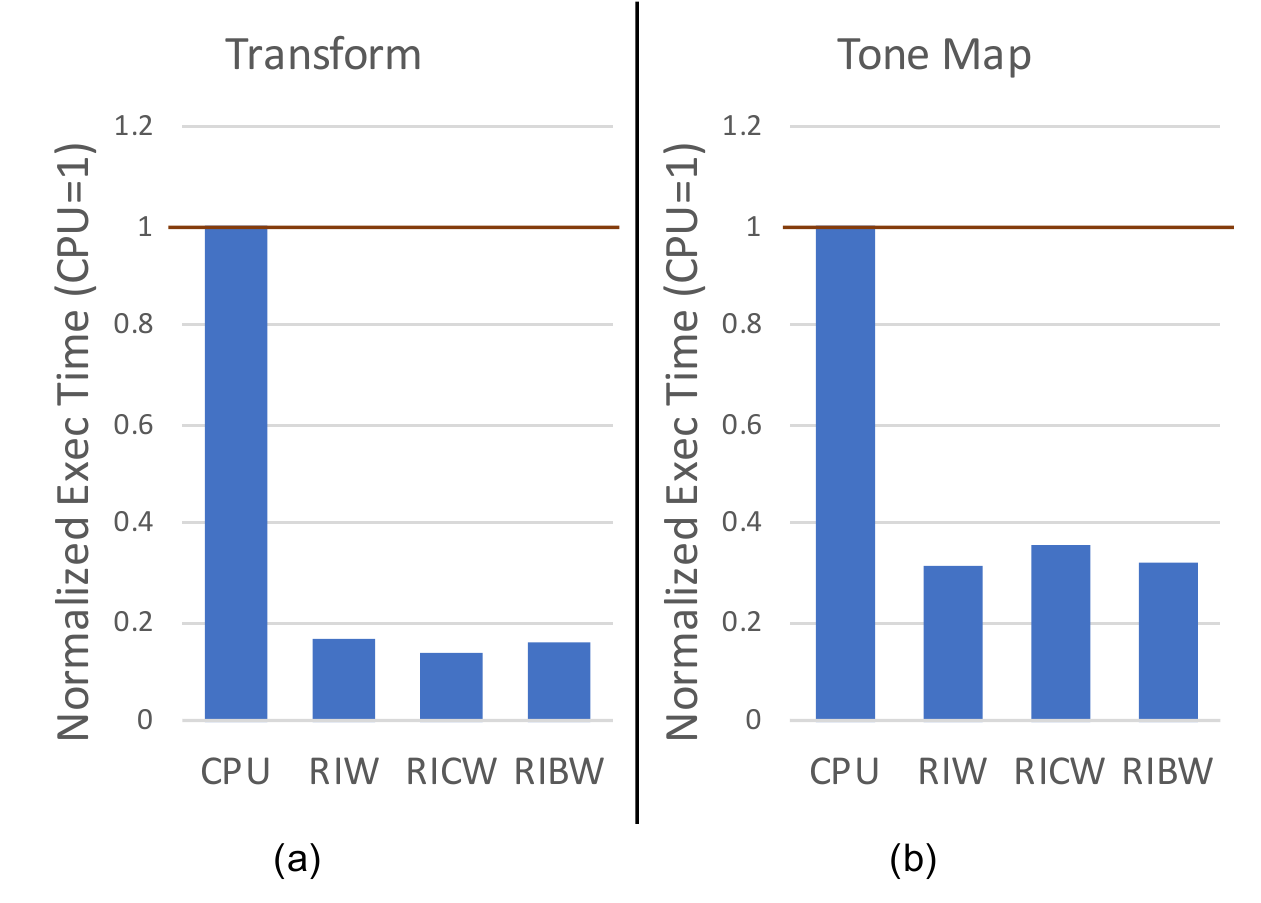}
    \caption{Effect of using constant memory (C) vs manual buffering (B) for read-only data. Does not apply to Denoise.}
    \label{fig:BC}
\end{figure}

\begin{figure}[h]
    \includegraphics[scale=0.6]{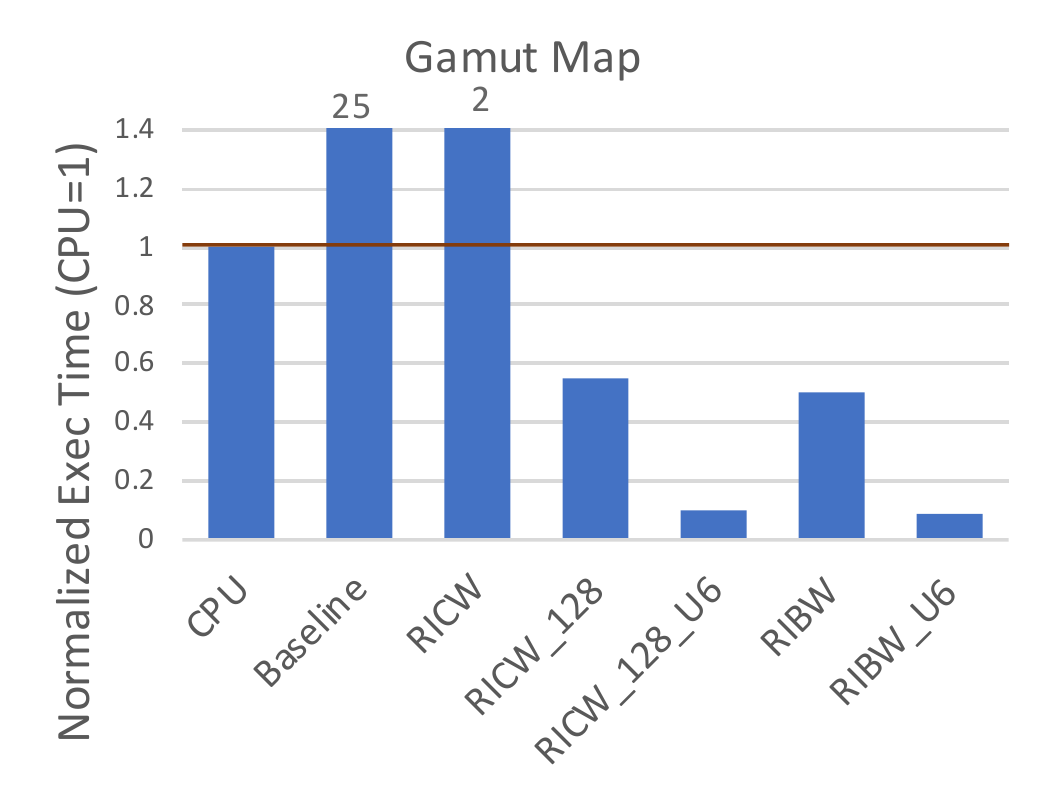}
    \caption{Different versions of Gamut Map kernel.}
    \label{fig:gamut}
\end{figure}

In this section, we provide an overview of the performance of the full pipeline, before diving into the different steps of our optimization strategy and analyzing each kernel separately. Figure \ref{fig:full} (a) presents the execution time for the different versions of the full pipeline, normalized to CPU. Two versions were tested: \textbf{Best:} has the best performing version of each kernel constituting the pipeline; \textbf{Auto:} has the best performing version of each kernel, with automatic optimizations only (as described in Section \ref{ssec:autoopt}), constituting the pipeline. Note that since the Baseline described in Section \ref{sec:opt} is very unoptimal, we do not use it as a comparison point in the overall performance study of the pipeline. The results in the graph clearly show that there is a big gap in the overall performance benefits of automatic optimizations that can be performed by a tool like AOC (3.3$\times$ slowdown over CPU), and the best performance possible using manual modifications of the code (21$\times$ speedup over CPU). 

We also study the effects of channels on the pipeline (Figure \ref{fig:full} (b)). We compare two versions against the best case from Figure \ref{fig:full} (a), one that uses constant memory with channels (C\_CH) and another that uses manual buffering with channels (B\_CH). We see that in our application Intel channels do not provide us with much more speedup compared to the best version of each kernel that uses global memory. The reason behind that is the imbalance between the kernels of the pipeline, which in this case causes increased stalling where the consumers are waiting on the producers to generate the data. We also show that manual buffering of the data gives better performance than using constant memory. This is due to the overhead in accessing the cache structure that is generated for constant memory compared to accessing the local memory that would be used for buffering.

\subsection{Single Kernel Results}
As discussed in Section \ref{sec:opt}, our optimization strategy involves four steps, applied incrementally to the baseline. We discuss the analysis of our strategy, one step at a time, in this section.

\subsubsection{Step 1:} This step studies the effect of the `\texttt{restrict}' (R) and `\texttt{ivdep}' (I) keywords on each kernel. Figure \ref{fig:RI} shows the three versions (R, I, and RI) compared to baseline and CPU. Although we also applied this step to Gamut Map, we do not show any results because these three versions of Gamut Map did not fit on our FPGA, and therefore we were not able to gather results for them. Our results show that there is always benefit in adding the two keywords, which is expected since this allows AOC to avoid making any assumptions related to memory aliasing and inter-loop memory dependence, thus allowing it to parallelize memory accesses to the fullest. Although RI isn't always the best-performing version, we chose it as the version to build on in our next steps in order to maintain uniformity accross all kernels. We also believe that R and I together will be beneficial for AOC to exploit more memory parallelism when we apply further optimizations.

\subsubsection{Step 2:} This step studies the effect of performing manual code modifications on top of RI, these include a combination of loop unrolling, interchange, and fusion, in order to increase the amount of parallel operations that can be scheduled in one iteration of the pipelined outerloop (as described in Section \ref{ssec:manopt} above). Our results are presented in Figure \ref{fig:W}, where W corresponds to manual rewrites. The results show that manual code modifications are crucial, as they can provide huge performance benefits (2.16$\times$-3.34$\times$ speedup vs RI) because of their ability to extract more independent operations per outerloop iteration, which can be spatially parallelized. Note that this optimization cannot be applied to Demosaic due to the nature of the kernel. Also, Gamut Map's version of RIW does not fit on our FPGA.

\subsubsection{Step 3:} This step studies the effect of further optimizing accesses to read-only memory after performing manual code modifications. This can either be done using constant memory (C) or manual buffering of the data in local memory (B). Figure \ref{fig:BC} and Figure \ref{fig:gamut} present the results for applying C and B on top of RIW. For Tone Map, our read only-data is larger and less frequently accessed, thus we believe that the overhead of the constant memory cache accesses slightly hurt performance. We also see that RIW performs slightly better than using either C or B, which we found is due to the fact that AOC was able to schedule it at a faster frequency compared to the other two. Another interesting observation is that with C and B added, the Gamut Map kernel now fits in our FPGA. We added an extra configuration to Gamut Map (\_128), which corresponds to manually configuring the size of the constant cache to 128KB, instead of the default 16KB, since the amount of read-only data accessed by this kernel is closer to 128KB (cache size can only be set to a power of 2). We see that by moving from a cache size that is smaller than the data set, to one that allows the data set to fit, greatly improves performance (\textasciitilde6$\times$ speedup vs \textasciitilde2$\times$ slowdown, when compared to CPU.). We also see that, although they are close, using buffering is slightly better than using constant memory. Finally, for Transform, the amount of read-only memory is small enough (9 floats), and the variation in the execution time is small enough, making the effect of C vs B negligible to the bigger picture. Therefore, the results show that it is almost always better to use buffering instead of constant memory when the data can fit on the FPGA, especially when the amount of data being accessed is significant.

\subsubsection{Step 4:} In addition to all the above optimizations, we perform an extra step for Gamut Map, which is unrolling its inner loop by a factor of 6. This optimization is not possible for the other kernels since they do not posses such an inner loop. We see in Figure \ref{fig:gamut} that by unrolling, we further improve the performance compared to buffering and using constant memory (\textasciitilde 36$\times$ vs CPU with unrolling, \textasciitilde 6$\times$ without). This is expected, since by unrolling the inner loop we are increasing the amount of independent operations that can benefit from the spatial parallelism on the FPGA. 


%% file: Related.tex
\section{Related Work} \label{sec:related}
One recent work that also analyzes AOC is presented in \cite{wang2016performance}. In this work the authors build an analysis framework for modeling the effects of different AOC optimization techniques. Their goal is to provide a tool for designers to be able to analyze the performance effects of difference AOC optimzations on their code. However, their work does not focus on determining the potential of AOC's automatic optimizations. Their work is that they focus on the multi-threaded execution model of AOC, rather than the Single Work Item model that we focus our study on because it is recommended by Intel to allow AOC to perform more automatic optimizaiton. Given that the multi-threaded model of AOC depends on the programmer to manually express the parallelism, most of the optimizations there require manual hand-tuning. 

There are recent surveys that focus on comparing HLS tools \cite{campbell2017new, nane2016survey}. These surveys provide a description of the abilities of different HLS tools, and compare ability of these tools in generating high-performing FPGA designs. However, they do not highlight what is possible with automatic optimizations in the different tools, versus what requires extensive hand-tuning.

Multiple academic papers present new HLS tool flows that can perform better automatic optimizations than AOC ~\cite{canis2013legup,canis2011legup,lee2016openacc,Zuo:2013:IPC:2555692.2555707,gupta2004coordinated,wei2013improving}. However, these papers do not specifically study the potential of automatic optimizaiton in AOC (or other commercial tools) compared to non-automatic optimizations. Instead, they show an overall improvement in generated code quality, and design efficiency. In addition, some of them are domain-specific targeting specific \cite{hegarty2014darkroom,pu2017programming,koeplinger2018spatial}. We focus on evaluating automating optimizations, and use AOC as a leading example of a commercial HLS tool that provides such optimizations.

%% file: Conclusion.tex
\section{Conclusion} \label{sec:conclusion}
In this paper, we presented a study of the automatic optimization potential of the AOC compiler, and the tradeoffs in using differnt optimization techniques. We show that there is limited potential in automatic optimziations that AOC can perform, even driven with programmer directives (2.8$\times$ speedup vs. CPU in the best case), and that great performance benefits can be achieved by combining them with other optimizations that require manual hand-tuning of code (36.5$\times$ speedup vs. CPU). This motivates the need for more extensive compiler optimizations in commercial HLS tools, an area of research that we are interested in. We also show that different optimizations have different effects on different kernels, and that while some have clear-cut effects, others depend on the behavior of the kernel.